\documentclass[apj,twocolumn,numberedappendix]{openjournal}

\usepackage{amsmath}
\usepackage{graphicx}
\usepackage{xcolor}
\usepackage{booktabs}
\usepackage{multirow}
\usepackage{adjustbox}
\usepackage{orcidlink}
\usepackage{float}
\usepackage{placeins} 

\hypersetup{
  colorlinks=true,
  citecolor=blue,
  linkcolor=blue,
  urlcolor=blue
}

\newcommand{\gaia}{\textit{Gaia }}

\begin{document}

\title{Revisiting Stellar equatorial rotational velocities with Gaia DR3 line 
broadening---the dependence on temperature, mass and age} 
\author{Amitay Sussholz\,\orcidlink{0009-0007-1546-1845}$^{1}$}
\author{Tsevi Mazeh\,\orcidlink{0000-0002-3569-3391}$^{1,2}$}
\author{Simchon Faigler\,\orcidlink{0000-0002-8368-5724}$^{1}$}

\affiliation{$^1$ School of Physics and Astronomy, Tel Aviv University, Tel Aviv, 6997801, Israel}
\affiliation{$^2$ Max-Planck-Institut für Astronomie, Königstuhl 17, D-69117 Heidelberg, Germany}

\email{amitays@mail.tau.ac.il}

\begin{abstract}

We used more than $10^5$  \gaia DR3 line broadening \texttt{vbroad} measurements to examine stellar rotation as a function of stellar temperature, mass and age.
%
%
The large sample clearly displays the Kraft break at $\sim 6{,}500$\,K, or mass of $\sim1.3\,M_{\odot}$---while \texttt{vbroad} are small, on the order of $10$--$20$ km/s, for stars cooler than the Kraft break, they sharply rise above the break, reaching up $\sim100$ km/s with temperature of $7{,}000$ K. To follow the stellar rotation as a function of age, we 
consider \texttt{vbroad} as a function of scaled age---stellar age divided by the relevant Terminal Age Main Sequence (MS), for four narrow mass bins. 
We find that stellar rotation deceleration is slow during the MS phase and fast afterwards for stars hotter than the break, whereas deceleration rate is relatively high and does not vary much for the cool stars.  
Our findings are consistent with the theory that stellar rotation slowing is due to magnetic breaking, emanating from magnetic fields that are anchored to the stellar convective envelopes. Therefore, deceleration is high in cool stars, but in hot stars only after they leave the MS and develop convective outer layers.

\end{abstract}

\section{Introduction}
\label{intro}

Stellar rotation is a fundamental property of stars, intimately linked to their internal structure, angular momentum evolution, and magnetic activity \citep{noyes84}. Rotation influences stellar winds, dynamo efficiency, internal mixing, and ultimately affects stellar lifetimes and planetary environments \citep{kraft67, kawaler87, bouvier97, wolff97}. Understanding how stars lose angular momentum across the main sequence (MS) and beyond has therefore remained a central question in stellar astrophysics for more than half a century \citep{weber67, mestel68, skumanich72, bouvier14, lu22, amaral25}.

The pioneering work of \citet{kraft67} established the sharp transition in rotation rates around spectral type F, now known as the \emph{Kraft break}. 
Cooler stars spin down efficiently through magnetized stellar winds,
whereas hotter stars (\(T_{\rm eff} \gtrsim 6{,}200\,{\rm K}\)) retain high rotation velocities due to their shallow convection zones that do not allow strong magnetic fields
\citep[e.g.,][]{kraft67, kawaler87, bouvier97, wolff97}.
Subsequent studies refined this picture, exploring the relation between angular momentum, mass, and stellar activity 
\citep{vansaders12, vansaders13b, mcquillan14, beyer24}.
Furthermore, some studies even suggest that the stellar rotation might be used as a measure of stellar age for stars below the Kraft brake \citep[e.g.,][]{skumanich72, barnes03, barnes07, angus15}.

With the advent of \textit{Kepler} and \textit{TESS}, tens of thousands of stellar rotation periods were measured from photometric variability \citep{nielsen13, mcquillan14, santos19, santos21, reinhold23, kamai25, gao25}. These studies revealed complex rotation-period distributions, strong mass dependencies, and evidence for weakened magnetic braking at old stellar ages \citep{vansaders13, matt15, vansaders16, avallone22}.

However, photometric methods inherently suffer from several drawbacks (see recent discussion by Binnenfeld et al., 2026, submitted). Rotation periods can only be detected when stars display surface inhomogeneities such as spots, which are often absent or evolve rapidly \citep{aigrain15}. Measured periods may also correspond to spot evolution timescales or harmonics rather than the true stellar rotation rate, introducing some biases \citep{reinhold13, mcquillan14}. Furthermore, old and slowly rotating stars tend to exhibit weak variability, leading to strong selection effects against long-period rotators [e.g.,][]\citep{metcalfe16}. These limitations motivate complementary approaches based on spectroscopy, where projected rotational velocities, derived from the broadening of the absorption lines, provide an independent tracer of stellar rotation, albeit with unknown stellar rotational inclination.

Admittedly, spectroscopic measurements can be affected by other sources of line broadening, most notably macroturbulence, which may bias the inferred rotational velocity \citep[e.g.,][]{Gray2005, Doyle2014}. Moreover, deriving stellar rotation periods from spectroscopic measurements requires knowledge of the stellar radius, introducing further uncertainties. Nevertheless, the \gaia revolution, that made millions of stellar spectra available, provides a powerful complementary avenue to photometric studies of stellar rotation.

The \gaia space mission carries a spectrometer with intermediate resolving power \(R \approx 11{,}500\), covering the 846--870~nm wavelength range, with the primary goal of measuring radial velocities (RVs) for bright sources \citep{cropper18, sartoretti22}, down to magnitude \(G_{\rm RVS} \sim 16\) \citep{katz23}. The spectroscopic pipeline \citep{sartoretti18} also derives a line-broadening parameter, estimating the broadening of absorption lines relative to template spectra, producing \texttt{vbroad} values for $3{,}524{,}677$ sources in \gaia DR3 \citep{fremat23}. 

This parameter, while not a direct measurement of projected rotational velocity (\(v \sin i\)), provides an important statistical tracer of stellar rotation. Although the relatively low resolution of the \gaia spectra limits the {\texttt{vbroad}} precision, the unprecedented large number of measurements opens new avenues of research \citep[e.g.,][]{hadad25a, hadad25b}, which was not possible before. This potential is further enhanced by the availability of stellar radii and masses from \textit{Gaia} \citep{gaia23, andrae23, creevey23} for this exceptionally large sample.


In this work, we used \gaia DR3 measurements to explore the distribution of projected rotation and specific angular momentum across the MS and beyond, with particular emphasis on stars below and above the Kraft break. We construct a color-magnitude 
diagram to exclude giant stars from our sample, and examine the dependence of \gaia \texttt{vbroad} on stellar mass and age.
Our analysis highlights the distinct rotational behaviors above and below the Kraft break and provides new empirical constraints on models of angular momentum evolution.

In Section~\ref{section:data} we describe the \gaia DR3 \texttt{vbroad} sample, the applied quality cuts, and the construction of a clean main-sequence (MS) subset, including our use of FLAME masses and ages and the age-scaling relative to an interpolated terminal-age MS. 
In Section~\ref{section:Velocities} we present the distribution of \texttt{vbroad} as a function of effective temperature and mass, reproducing the break seen in Kraft's original paper. 
In Section~\ref{section:age} we examine the evolution of \texttt{vbroad} and the derived specific angular momentum as a function of scaled age in narrow mass bins, contrasting stars above and below the Kraft break. 
Finally, we discuss the implications of our results and dominant sources of uncertainty in Section~\ref{section:discussion}.

\section{Data Selection}
\label{section:data}
We base our analysis on  $3{,}524{,}677$ line-broadening measurements (\texttt{vbroad}) published in \gaia Data Release 3 \citep{fremat23}.
%
%
We used ADQL queries via the Table Access Protocol (TAP)\footnote{\url{https://www.star.bris.ac.uk/mbt/topcat/sun253/TapTableLoadDialog.html}} service \citep{salgado17} to retrieve data from the \gaia DR3 source catalog \citep{gaia23}. Stellar parameters from GSP-Phot (e.g., \texttt{teff\_gspphot}, \texttt{radius\_gspphot}) were taken from \citet{andrae23}, and evolutionary parameters from FLAME (e.g. \texttt{mass\_flame}, \texttt{age\_flame}) were taken from the CU8/Apsis DR3 products \citep{creevey23}.

To ensure reliable and astrophysically meaningful measurements, we applied the following quality cuts:

\begin{itemize}
    \item \( \texttt{vbroad\_error} < max(10,0.5*\texttt{vbroad})\): we require a detection significance of at least \(2\sigma\) to exclude poorly constrained broadening measurements, which could otherwise bias the inferred rotation distribution, but allowing for near zero values.
    \item \(4{,}000\,{\rm K} < T_{\rm eff,\,GSPphot} < 8{,}000\,{\rm K}\): this range selects FGK MS stars, covering the cool dwarf regime and extending up to early F-type stars above the Kraft break.
    \item \(\varpi / \sigma_{\varpi} > 10\): accurate distances are essential to compute reliable absolute magnitudes and radii. This cut removes stars with poorly constrained parallaxes, minimizing contamination in the CMD \citep{gaia18}.
    \item \(R_{\rm GSPphot} / \sigma_{R,\,{\rm lower}} > 5\): we retain only stars with well-determined radii, ensuring that subsequent calculations of angular momentum are robust.
    \item \(M_{\rm FLAME} / \sigma_{M,\,{\rm lower}} > 5\): similarly, we require well-determined stellar masses from the \gaia FLAME module.
\end{itemize}
After these cuts, our working sample contained $1{,}300{,}637$ stars, of which $1{,}116{,}716$ have an age value in FLAME.

To further restrict the sample to MS stars, we constructed a color--magnitude diagram (CMD) using extinction-corrected \gaia photometry. To isolate MS stars, including mildly evolved ones, we defined a scaled evolutionary age as the ratio of the FLAME age to the corresponding terminal-age main sequence (TAMS), $t_{\rm FLAME}/TAMS$. The TAMS values were defined following \citet{vansaders13b} as the age at which the central hydrogen abundance satisfies $\mathrm{\Large\chi}_c<2 \times 10^{-4}$.

To compute TAMS as a function of stellar mass, we downloaded the \texttt{"fastlaunch\_full.tar.gz"} and \texttt{"slowlaunch\_full.tar.gz"} grids from \citet{claytor25} using the \texttt{kiauhoku}\footnote{\url{https://zenodo.org/records/14908017}} package
 \citep{claytor20} based on YREC models at solar metallicity. These grids were used to construct a lookup table of stellar masses and their corresponding TAMS values. For stars with masses in the range $0.68 \le M/M_\odot \le 2$, we interpolated the TAMS using a Piecewise Cubic Hermite Interpolating Polynomial (PCHIP)\footnote{\url{https://www.mathworks.com/help/matlab/ref/pchip.html}}. For masses outside this range, we extrapolated the TAMS values using a power-law fit.

\begin{figure}[ht]
    \centering
    \includegraphics[width=0.95\columnwidth]{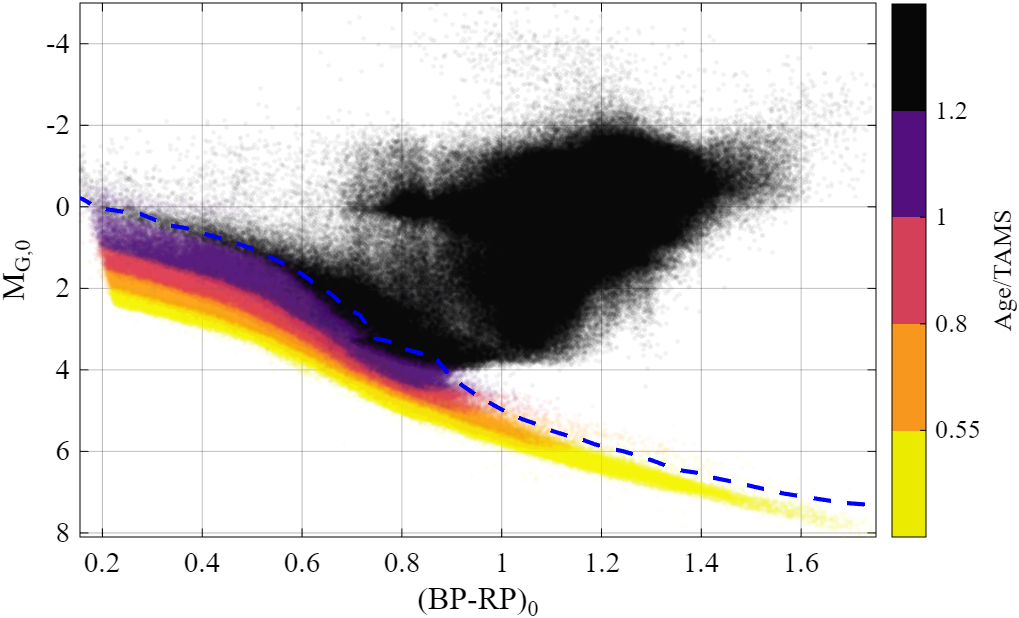}
    \caption{
        Extinction-corrected \gaia CMD of $1{,}116{,}716$ stars sample with reliable \texttt{vbroad} measurements from \gaia DR3, and an Age value in FLAME. 
        Point density is presented via point alpha value, colored by the stars' evolutionary stage ($Age/TAMS$) in 5 discrete ranges. 
        The dashed blue curve was used to separate the MS from evolved stars. 
        Stars below the curve are the final clean MS sample used in this work ($424{,}270$ stars).
    }
    \label{fig:Gaia_CMD}
\end{figure}

To avoid over culling our data and account for inaccuracies in the values of FLAME's ages, we decided to include slightly evolved stars with $t_{FLAME}/TAMS \ \leq1.2$. The final sample selection was performed through visual inspection of the stellar locations in the CMD, using $t_{FLAME}/TAMS$ as a color-coded guide, as illustrated in Figure~\ref{fig:Gaia_CMD}.
%
%
This cut effectively removes evolved stars, yielding a MS sample of $424{,}270$ stars for our subsequent analysis of stellar rotation and angular momentum. Focusing on the main bulk of our MS sample, we further restrict the data to stars with $0.8 \le M/M_\odot \le 1.7$ and $4{,}800 \,K \le T_{eff} \le 7{,}800 \,K$, resulting in a final sample of $354{,}849$ stars.

\section{\texttt{vbroad} as a function of temperature}
\label{section:Velocities}

\begin{figure}[th!]
    \centering
    \includegraphics[width=0.95\columnwidth]{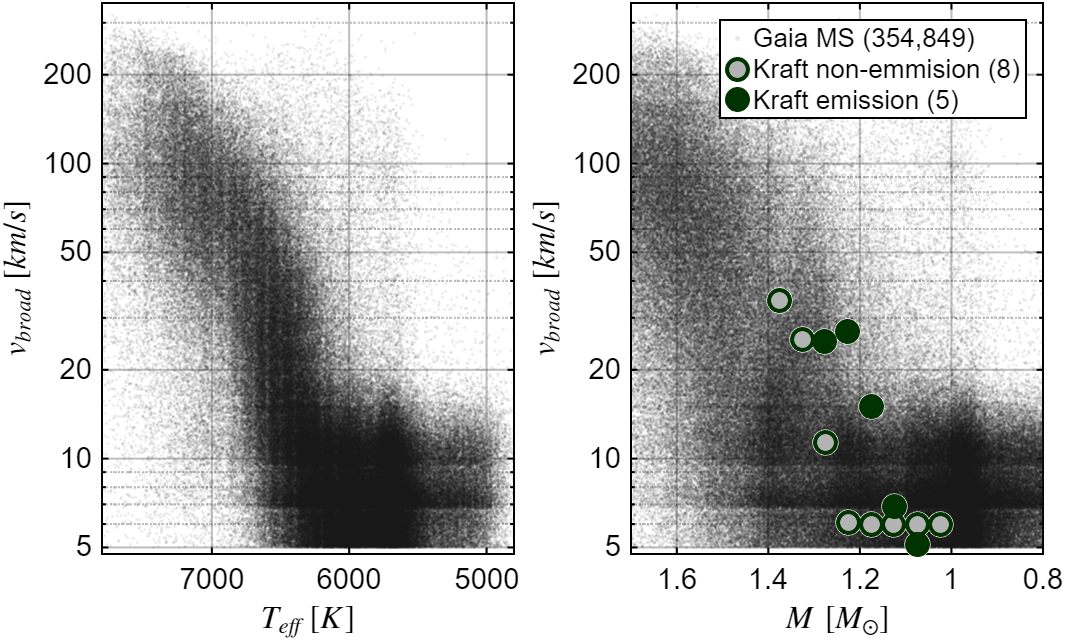}
    \caption{{\it Left}: Projected rotational velocity (\texttt{vbroad}) versus effective temperature, taken from \gaia GSPphot. Hotter stars tend to rotate more rapidly as presented here.
    {\it Right}: Projected rotational velocity (\texttt{vbroad}) versus stellar mass, taken from \gaia FLAME for $354{,}849$ \gaia MS stars. Filled and open circles are data points taken from Fig 3. of \cite{kraft67}. }
    \label{fig:Gaia_vsini_teff_Mass}
\end{figure}

 Figure~\ref{fig:Gaia_vsini_teff_Mass} displays \texttt{vbroad} values as a function of stellar effective temperature and mass.
The Kraft break is clearly seen. For 
$T_{\rm eff} \lesssim 6{,}250$ K ($M\lesssim 1,3$--$1,4 M_{\odot}$) the rotational velocities are mostly less than $\sim10$ km/s, which is below the resolution of \gaia RVS. Beyond this range, we see a strong rise of \texttt{vbroad} as a function of temperature or mass.


We overplotted the thirteen measurements of \citet{kraft67} on the figure. 
Remarkably, the trend identified by Kraft, based on only 13 stars, is fully confirmed by the \textit{Gaia} \texttt{vbroad} measurements of several hundred thousand stars.

\section{\texttt{vbroad} and specific angular momentum as a function of age}
\label{section:age}

\begin{figure*}
    \centering
    \includegraphics[width=0.95\textwidth]{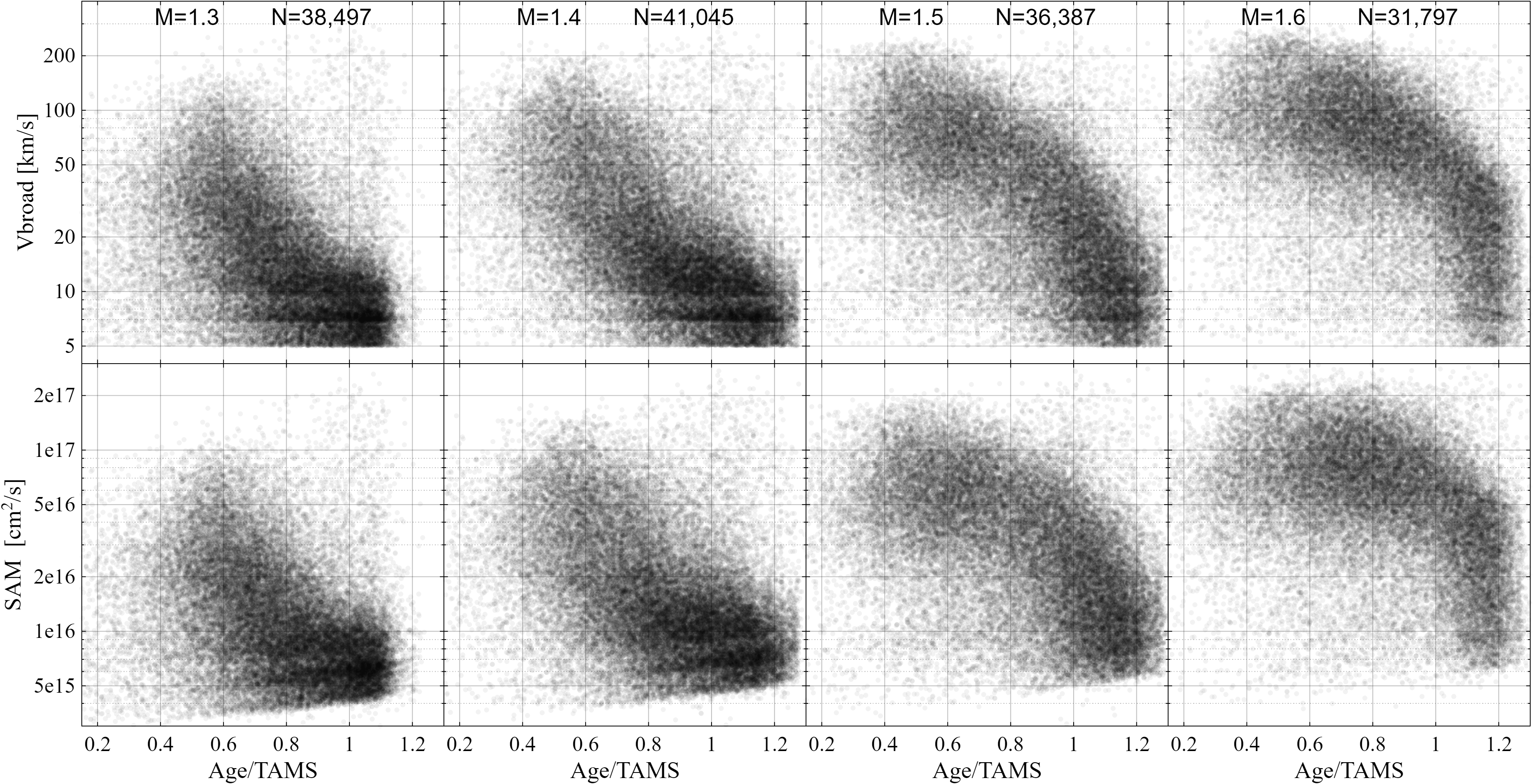}
    \caption{
    {\it Upper:} Stellar Gaia's \texttt{vbroad} vs.~the corresponding GSP-Phot's evolutionary stage in four FLAME-mass bins. with width of  $0.1\,M_\odot$. Bin centers are at $1.3\,M_\odot$, $1.4\,M_\odot$, $1.5\,M_\odot$ and $1.6\,M_\odot$. 
%
We use \gaia scaled age---defined as the stellar age in units of the stellar interpolated TAMS 
(see text).
Our distributions are plotted as translucent black dots, signifying point density.
{\it Lower}: The stellar specific angular momentum vs.~its scaled age.
}
    \label{fig:Gaia_AgeV_Slice}
\end{figure*}

Figure~\ref{fig:Gaia_vsini_teff_Mass} shows a pronounced dependence of \texttt{vbroad} on stellar mass and effective temperature with a large scatter, likely driven by the wide range of stellar ages in the sample.
To study the dependence on stellar age, we plot in Figure~\ref{fig:Gaia_AgeV_Slice} \texttt{vbroad} as a function of stellar age for four slices of masses. 

We scaled the age by the TAMS, so age of unity corresponds to the transition of the star from its MS phase to the relatively fast expansion phase. By construction, the figure extends up to a scaled age of $1.25$, allowing the rotational-velocity evolution to be traced beyond the end of the MS phase.

In order to follow the stellar angular momentum loss we plotted  in the lower panel of the figure the stellar specific angular momentum (SAM) as a function of their age, with the same mass bins. The SAM of each star was derived by multiplying its equatorial rotational velocity \texttt{vbroad} by its FLAME radius. 

Figure \ref{fig:Gaia_AgeV_Slice} displays a significant difference in 
the age-dependence pattern of \texttt{vbroad} and SAM for stars below and above the Kraft break. 
Stars above the Kraft break, with $1.5$ and $1.6 M_{\odot}$, display a slow rate of deceleration of their rotational velocity and their SAM at their MS phase, and then change into a much faster rate as they evolve off the main sequence. In contrast, stars with masses below the Kraft break show opposite behavior---fast deceleration of stellar rotation and SAM during the main sequence and much slower deceleration at the post-MS phase.

\section{Discussion}
\label{section:discussion}

We used more than $10^5$  \gaia \texttt{vbroad} measurements to examine the deceleration of stellar rotation as stars evolve along and beyond the main sequence. 
The analysis is affected by two main sources of uncertainty. 
The first concerns the derivation of stellar equatorial rotational velocities. 
The medium spectral resolution of \gaia introduces relatively large \texttt{vbroad} uncertainties, of the order of $10$ km/s, such that values below $\sim 20$ km/s are likely not robust. 
The \texttt{vbroad} measurements are also affected by additional  broadening effects, such as macro-turbulence, which tend to increase the inferred values \citep{fremat23},  while the unknown inclination leads to an overall underestimation of the true equatorial velocity \citep[e.g.,][]{hadad25a, hadad25b}. 
Finally, the 
limited wavelength coverage of the RVS spectra and their center in the red part of the spectrum limit the ability to derive equatorial rotational velocities.

The second source of uncertainty is associated with the stellar 
parameters---effective temperature, radius, mass, and age, adopted from the FLAME catalog \citep{creevey23}. While the stellar radius and effective temperature are derived directly from \gaia photometry and parallax \citep{gaia23}, the stellar mass---and especially the 
age---are strongly model dependent \citep{fouesneau23}.
Despite these limitations, the large size of the sample allows clear and statistically robust patterns of stellar rotation evolution to emerge.

We therefore focus on the equatorial rotational velocities and do not consider the stellar rotation periods, in order to avoid the need to divide the \texttt{vbroad} values by the stellar radii, which would introduce an additional uncertainty. Nevertheless, when considering the specific angular momentum, we had no choice but to multiply \texttt{vbroad} by the stellar radius.

The large sample clearly reveals the Kraft break at $\sim 6{,}500$\,K, 
corresponding to a mass of $\sim1.3\,M_{\odot}$. Stars cooler than the Kraft break exhibit low \texttt{vbroad}, 
typically of order $10$--$20$ km/s. 
For stars above the break, equatorial velocities rise steeply as a function of temperature, up to $\sim100$ km/s for stars with $T_{\rm eff}\sim 7{,}000$K. 

The dependence of \texttt{vbroad} on stellar mass and effective temperature displays a large scatter, likely driven by the wide range of stellar ages in the sample.
To trace stellar rotation as a function of age, we examine \texttt{vbroad} in four narrow mass bins, spanning both sides of the Kraft break.
For each star, we used the scaled age—--defined as the stellar age divided by the corresponding terminal-age MS (TAMS) age \citep[see a similar approach in][]{sun24}.
We find distinctly different patterns of equatorial-velocity variability for stars above and below the break. 
For stars hotter than the break, stellar rotation deceleration is slow during the MS phase, 
but increases rapidly after MS departure.
For cooler stars below the break, the deceleration rate is relatively high and varies only weakly with evolutionary stage.

We also present the stellar specific angular momentum as a function of the scaled age, 
derived from the observed equatorial rotational velocities and the stellar radii. 
However, a cautionary note is due here. 
Although the derivation 
accounts for variations in stellar radius, it assumed a constant radius of gyration for all ages and for all stars, which is clearly an oversimplification.
This approximation may be adequate during the MS phase, but it becomes increasingly unrealistic after MS departure, when rapid stellar expansion dramatically alters the internal density profile.
Nevertheless, the derived SAM follows the same qualitative trends as the equatorial velocities, indicating rapid angular-momentum loss whenever the stellar envelope is convective.
The detailed evolution of angular-momentum loss after MS departure remains to be explored.

Our findings are consistent with the theoretical picture in which stellar spin-down is driven by magnetic braking associated with magnetic fields anchored in convective envelopes \citep[see the recent review by][]{aerts25}.
Consequently, rotational deceleration is strong in cool stars with deep convective envelopes, and in hotter stars only after they leave the MS and develop convective outer layers.
However, a detailed quantitative confrontation between theory and the statistical properties of stellar rotation still awaits larger and more homogeneous datasets. 

Gaia spectra are obviously not the only source for deriving stellar rotation. The LAMOST database \citep{LiuLAMOST20}, for example, has been used by \citet{sun24} to derive stellar parameters, including rotation periods, for $\sim10^5$ stars.
Another major multi-object project is GALAH \citep{Buder21,buder25}, which has yielded nearly a million rotational velocities \citep[e.g.,][]{das25}.
In the near future, 
ongoing and upcoming multi-object spectroscopic surveys will continue to produce large numbers of high-quality spectra,  including the upcoming 4MOST project \citep{4MOSTdejong19, 4MOST22}, and \gaia DR4,\footnote{https://www.cosmos.esa.int/web/gaia/data-release-4} expected at the end of 2026, which will release many more \texttt{vbroad} determinations.

The complementary approach of deriving stellar rotation periods from photometric variability will also provide a wealth of additional rotation measurements. The \textit{Gaia} rotation module \citep{distefano23} is expected to deliver a large number of stellar rotation periods based on \textit{Gaia} photometry. Within the next few years, the \textit{PLATO} space mission \citep{PLATO25} will further expand this effort by producing many stellar light curves that enable robust rotation-period determinations.

A comprehensive study that integrates all available datasets into a coherent picture of stellar rotation as a function of stellar 
parameters---especially age---is still needed.
This includes stellar rotation of stars in open clusters 
\citep[e.g.,][]{prosser95,meibom05,meibom15, fritzewski21}, where we know the stellar age quite well, and extension of the analysis to early-type \citep[e.g.,][]{ramrez15, blex24} and metal-poor stars
\citep{mokiem06}. 
Such a study will also need to confirm or refute unexpected features, such as the non-monotonic behavior of A-star rotational velocities reported by \citet{sun24}, which is not observed in our analysis.
\cite{sun24} also found a stellar rotation metallicity dependence, which should be followed carefully for stars of different masses.

A comprehensive picture of stellar rotational evolution will ultimately allow a quantitative determination of angular-momentum loss rates and their dependence on stellar properties \citep[see the seminal work of][]{mestel68}.
Ultimately, this will enable refinement of the gyrochronology techniques \citep{barnes07,barnes10,aerts25}, allowing more accurate age determinations based on stellar mass and rotation.


\bibliographystyle{aasjournal}
\bibliography{main}



\end{document}